\documentclass[12pt, draftclsnofoot, onecolumn]{IEEEtran}
\usepackage{cite}
\usepackage{graphicx,amsmath,amssymb}
\usepackage{subfigure}
\usepackage{citesort}
\usepackage{fancyhdr}
\usepackage{mdwmath}
\usepackage{mdwtab}
\usepackage{balance}
\usepackage{xcolor}
\usepackage{bm}

\usepackage{algorithm}
\usepackage{algorithmic}
\usepackage{multirow}
\usepackage{flafter}
\usepackage{comment}
\usepackage{flushend}
\usepackage{amsthm}

\newtheorem{theorem}{Theorem}

\newtheorem{corollary}{\textbf{Corollary}}

\begin{document}

\title{Non-Reciprocal Reconfigurable Intelligent Surfaces}

{

 \author{Jiaqi\ Xu, Haoyu\ Wang, Rang\ Liu, Josef A. Nossek,~\IEEEmembership{Life Fellow,~IEEE}, A. Lee Swindlehurst,~\IEEEmembership{Fellow,~IEEE}

\thanks{J. Xu, H.Wang, R. Liu, and A. Swindlehurst are with the Center for Pervasive Communications and Computing, University of California, Irvine, 92697, CA, USA. (email:\{xu.jiaqi, haoyuw30, rangl2, swindle\}@uci.edu).}
\thanks{J. Nossek is with the School of Computation, Information and Technology, Technical University of Munich, 80333 Munich, Germany. (email: josef.a.nossek@tum.de)}
}
\maketitle
\begin{abstract}
In contrast to conventional RIS, the scattering matrix of a non-reciprocal RIS (NR-RIS) is non-symmetric, leading to differences in the uplink and the downlink components of NR-RIS cascaded channels. In this paper, a physically-consistent device model is proposed in which an NR-RIS is composed of multiple groups of two-port elements inter-connected by non-reciprocal devices. The resulting non-reciprocal scattering matrix is derived for various cases including two-element groups connected with isolators or gyrators, and general three-element groups connected via circulators. Signal models are given for NR-RIS operating in either reflecting-only or simultaneously transmitting and reflecting modes. The problem of NR-RIS design for non-reciprocal beamsteering is formulated for three-element circulator implementations, and numerical results confirm that non-reciprocal beamsteering can be achieved with minimal sidelobe power. We also show that our physically consistent NR-RIS architecture is effective in implementing channel reciprocity attacks, achieving similar performance to that with idealized NR-RIS models.
\end{abstract}

\begin{IEEEkeywords} 
Channel non-reciprocity, device modelling, multi-port network, reconfigurable intelligent surfaces (RISs)
\end{IEEEkeywords}

\section{Introduction}

In recent years, reconfigurable intelligent surfaces (RIS) have been proposed to assist wireless communication networks through purposeful control of the radio environment~\cite{di2019smart}. RIS are two-dimensional planar structures that consist of a large number of low-cost elements of sub-wavelength sizes whose electromagnetic responses can be adaptively adjusted. 
Recently, several variants of conventional RIS have been introduced to further push the limit of this technology, including simultaneously transmitting and reflecting (STAR)-RIS~\cite{xu_star}, beyond diagonal (BD)-RIS~\cite{9913356}, and active RIS~\cite{9998527}. However, existing research contributions almost exclusively focus on RIS that are reciprocal, i.e., RIS with symmetric phase-shift or scattering matrices. 
In~\cite{9690479}, it was shown that an RIS-aided channel is reciprocal if the permittivity and permeability tensors of the RIS are symmetrical. The authors of~\cite{10201815} developed a path loss model for non-reciprocal (NR)-RIS assuming the existence of two independent diagonal phase-shift matrices for the uplink and downlink signals. However, no discussion of how to practically obtain this non-reciprocity was discussed.

In the domain of physics, however, pioneering research contributions have already demonstrated non-reciprocity using time-modulated gradient~\cite{PhysRevB.92.100304,PhysRevApplied.11.054054}
or modulated metasurfaces~\cite{Zhang}.
Nevertheless, these contributions have not found their way into wireless systems engineering since they are typically focused only on beamsteering or they produce significant frequency shifts between the incident and reflected waves. In addition, there is a lack of physically-consistent device models that would enable the design and optimization of NR-RIS in wireless applications.

\begin{figure}[t!]
    \begin{center}
        \includegraphics[scale=0.3]{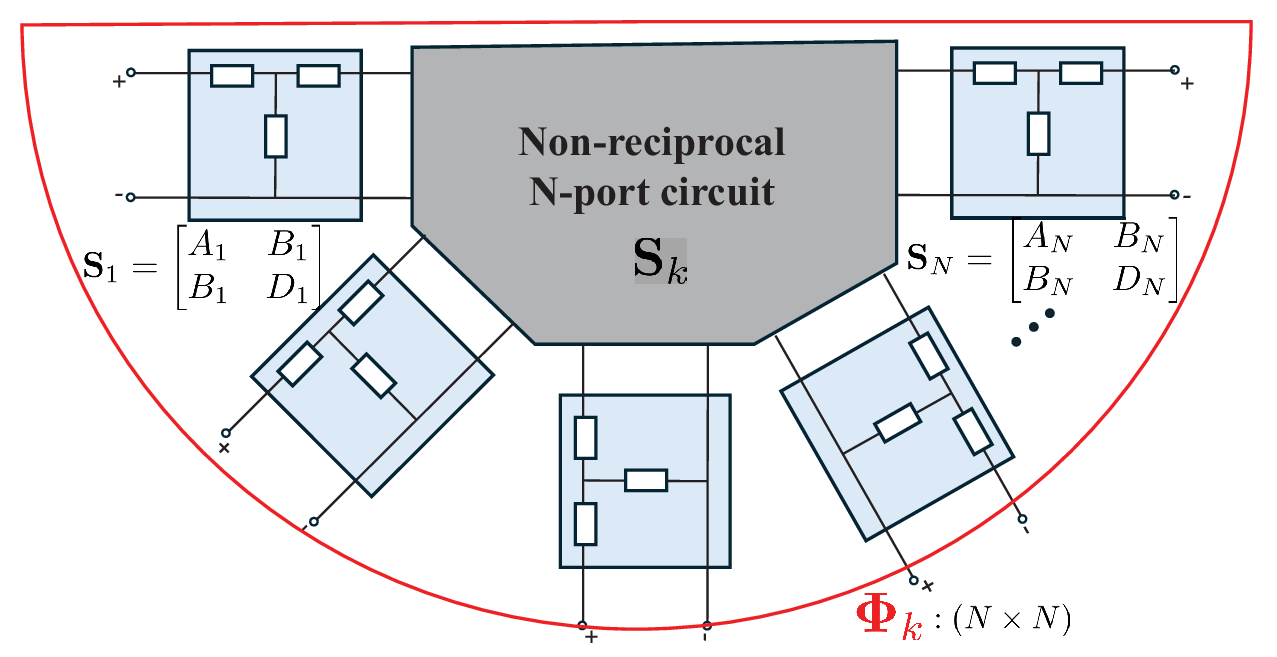}
        \vspace{-0.1 cm}
        \caption{Schematic illustration of an NR-RIS connected $N$-element group.}
        \label{Fig1}
    \end{center}\vspace{-0.2 cm}
\end{figure}

To bridge this gap, in this paper we propose a physically-consistent device model for non-reciprocal (NR)-RIS based on multiport communication theory~\cite{5446312}. In particular, we consider a BD-RIS-type architecture in which the surface elements are inter-connected and allow inter-element signal transfer. However unlike the reciprocal BD-RIS approach, we assume that non-reciprocal devices are used to connect the elements, resulting in a non-symmetric overall scattering matrix. An early speculative version of the BD-RIS architecture posited that a surface could achieve a non-symmetric scattering matrix by channeling all energy incident on one element and transmitting it from another \cite{9737373}. This NR architecture was exploited as a means for launching channel reciprocity attacks in~\cite{10445725}.
However, we will see that our physically consistent model does not allow for such a configuration, hence the importance of considering realizable circuit models for RIS in general.

In Section. II, we first present a two-port model for each RIS element and propose three practical realizations for NR-RIS using isolators, gyrators, and circulators. Then, in Section. III, we provide signal models for NR-RIS operating in both reflecting-only and STAR modes. Based on the proposed models, we demonstrate how NR-RIS can achieve reconfigurable non-reciprocal beamsteering. Finally, some numerical results are provided in Section V.

\section{Physically Consistent NR-RIS Models}\label{sec_hardware}

We consider an $N$-element NR-RIS divided into $K$ groups, where the $k$-th group is composed of $N_k$ elements inter-connected with an NR device, and $\sum_{k=1}^K N_k = N$.
A schematic representation of one inter-connected group of NR-RIS elements is shown in Fig.~\ref{Fig1}.
Each individual RIS element is modeled as a two-port reciprocal network where one end of the element interacts with free space and the other is connected to an NR $N_k$-port circuit. The NR circuit has a non-symmetric scattering matrix $\mathbf{S}_k$, and we let $\mathbf{\Phi}_k$ denote the scattering matrix (``$S$-matrix'') of the $k$-th group. Then, the overall scattering matrix of the NR-RIS can be expressed as a block-diagonal matrix: $\bar{\mathbf{\Phi}} = \mathrm{blkdiag}\{\mathbf{\Phi}_1,\mathbf{\Phi}_2,\cdots,\mathbf{\Phi}_K\}$.

The $n$-th RIS element is characterized by a $2\times 2$ reciprocal impedance matrix:
\begin{equation}\label{Zn}
    \mathbf{Z}_n = \begin{bmatrix}
Z_{11} & Z_{12} \\
Z_{12} & Z_{22}
\end{bmatrix}.
\end{equation}
As shown in Fig.~\ref{Fig1}, 
the scattering matrix of the $n$-th element is given by~\cite{pozar2011microwave}:
\begin{equation}\label{abcd}
    \mathbf{S}_n = \begin{bmatrix}
A_n & B_n \\
B_n & D_n
\end{bmatrix} = (\mathbf{Z}_n-\mathbf{I}Z_0)(\mathbf{Z}_n+\mathbf{I}Z_0)^{-1} ,
\end{equation}
where $Z_0$ is the free-space impedance.

\subsection{Two-Element Group with Isolator}
\begin{figure}[t!]
    \begin{center}
        \includegraphics[scale=0.45]{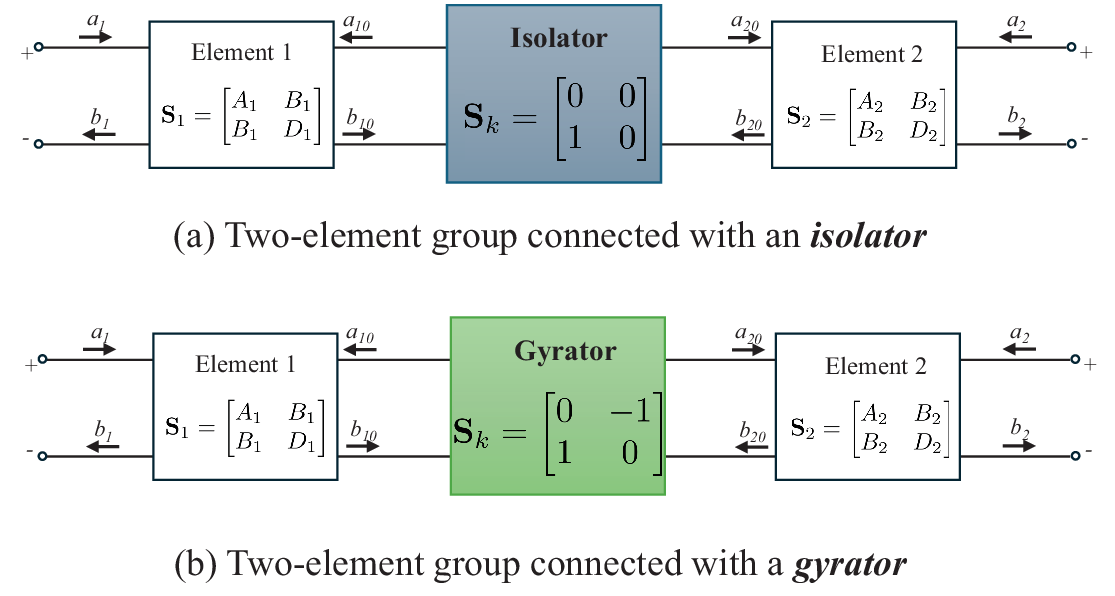}
        \vspace{-0.15 cm}
        \caption{Schematics of two-element NR-RIS groups.}
        \label{Fig2}
    \end{center}\vspace{-0.3 cm}
\end{figure}
An isolator is a two-port device that allows radio frequency (RF) signals to pass in one direction only. The non-symmetric scattering matrix of an ideal isolator is given by:
\begin{equation}\label{s_iso}
    \mathbf{S}_{\mathrm{iso}} = \begin{bmatrix}
0 & 0 \\
1 & 0
\end{bmatrix}.
\end{equation}
Fig.~\ref{Fig2} shows two RIS elements with scattering matrices $\mathbf{S}_1$ and $\mathbf{S}_2$ connected through an isolator. The following theorem gives the resulting scattering matrix of this two-element group.

\begin{theorem}
The non-symmetric scattering matrix of a two-element group connected by an isolator is given by:
\begin{align}\label{ISO_overll}
\mathbf{\Phi} = \begin{bmatrix}
A_1 & 0 \\
B_1B_2 & A_2
\end{bmatrix}.
\end{align}

\begin{proof}
From Fig.~\ref{Fig2}(a) we have
\begin{equation}\label{b1b10}
 \begin{bmatrix}
     b_1\\b_{10}
 \end{bmatrix}  = \begin{bmatrix}
A_1 & B_1 \\
B_1 & D_1
\end{bmatrix} \begin{bmatrix}
     a_1\\a_{10}
 \end{bmatrix}.
\end{equation}
The isolator in \eqref{s_iso} dictates that $a_{10} = 0$ and $a_{20} = b_{10}$. Inserting these into \eqref{b1b10}, we have $b_1 = A_1a_1$ and $a_{20}=b_{10} = B_1a_1$. Finally, using the scattering property of element 2, we have $b_2 = A_2a_2 + B_2a_{20}= A_2a_2 + B_2B_1a_1$. Rewriting this in matrix form, Theorem 1 is proved.

\end{proof}
\end{theorem}

\subsection{Two-Element Group with Gyrator}

A gyrator is a passive-lossless two-port that allows radio frequency signals to be transmitted in both directions, but one with a phase shift of $\pi$ relative to the other. The non-symmetric scattering matrix of an ideal gyrator is given by:
\begin{equation}\label{s_gy}
    \mathbf{S}_{\mathrm{gyra}} = \begin{bmatrix}
0 & -1 \\
1 & 0
\end{bmatrix}.
\end{equation}
Fig.~\ref{Fig2} shows two RIS elements with scattering matrices $\mathbf{S}_1$ and $\mathbf{S}_2$ connected through a gyrator, and Theorem~\ref{thm2} provides the corresponding scattering matrix.

\begin{theorem}\label{thm2}
The non-symmetric scattering matrix of a two-element group connected by a gyrator is given by:
\begin{align}\label{Gy_overall}
\mathbf{\Phi} = \begin{bmatrix}
A_1-B^2_1D_2/\Delta & -B_1B_2/\Delta \\
B_1B_2/\Delta & A_2-B^2_2D_1/\Delta,
\end{bmatrix},
\end{align}
where $\Delta = 1+D_1D_2$.
\begin{proof}
The proof is similar to Theorem 1 and thus omitted.
\end{proof}
\end{theorem}

\begin{figure}[t!]
    \begin{center}
        \includegraphics[scale=0.45]{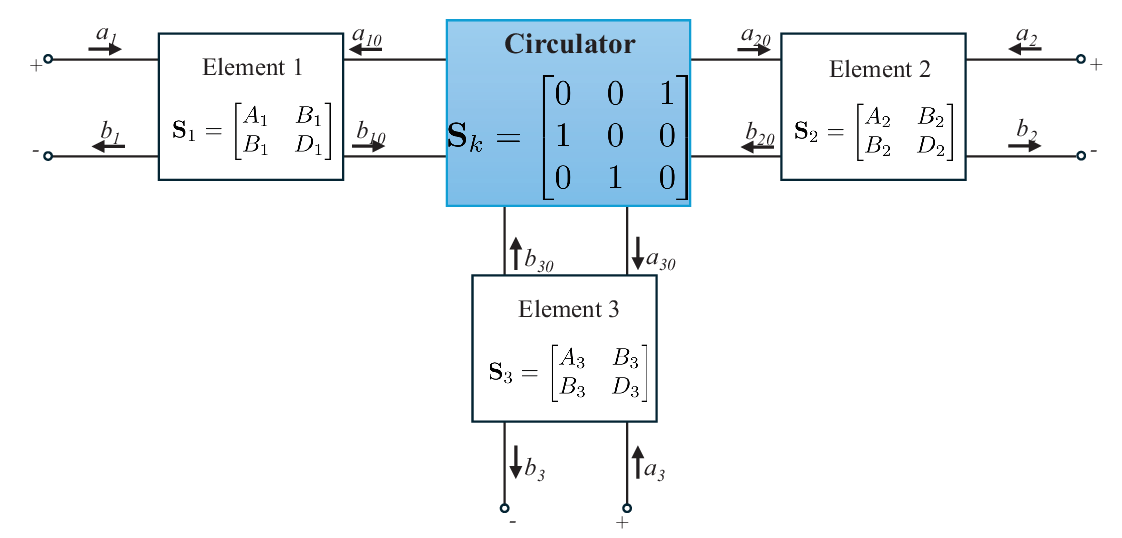}
        \vspace{-0.15 cm}
        \caption{A three-element NR-RIS connected via a circulator.}
        \label{Fig3}
    \end{center}\vspace{-0.3 cm}
\end{figure}

\subsection{Three-Element Group with Circulator}
A circulator is a passive three-port device that circulates RF energy from one port to an adjacent port. The non-symmetric scattering matrix for one type of ideal circulator is
\begin{equation}\label{s_cir}
    \mathbf{S}_{\mathrm{circ}} = \begin{bmatrix}
0 & 0 & 1\\
1 & 0 & 0\\
0 & 1 & 0
\end{bmatrix}.
\end{equation}
Fig.~\ref{Fig3} shows three RIS elements with scattering matrices $\mathbf{S}_1$, $\mathbf{S}_2$, and $\mathbf{S}_3$ connected through a circulator, producing the general $3\times 3$ scattering matrix derived in the following theorem.

\begin{theorem}
The non-symmetric scattering matrix of a three-element group connected by the circulator in Fig.~\ref{Fig3} is given by

\begin{footnotesize}
\begin{equation}\label{Gir_overall}
\mathbf{\Phi} = \begin{bmatrix}
A_1 + B^2_1D_2D_3/\Delta & B_1B_2D_3/\Delta & B_1B_3/\Delta \\
B_1B_2/\Delta & A_2+ B^2_2D_1D_3/\Delta & B_2B_3D_1/\Delta \\
B_1B_3D_2/\Delta & B_2B_3/\Delta & A_3 + B^2_3D_1D_2/\Delta,
\end{bmatrix},
\end{equation}
\end{footnotesize}
\noindent where $\Delta = 1-D_1D_2D_3$.
\begin{proof}
We provide the proof for element $[\mathbf{\Phi}]_{11}$; the other elements of $\mathbf{\Phi}$ can be found using similar steps. The elements of the scattering matrix of a general multiport are defined as:
\begin{equation}
    [\mathbf{\Phi}]_{mn} = \Big(\frac{b_m}{a_n}\Big ) \bigg\rvert_{a_k=0, k\ne n}.
\end{equation}
Using the scattering matrix \eqref{s_cir}, we have $b_{30}=a_{10}$, $b_{10}=a_{20}$, and $b_{20}=a_{30}$. Combining the scattering relations for all three elements, we can obtain:
\begin{equation}
   a_{10} =b_{30} = D_3a_{30} = D_3b_{20} = D_3D_2b_{10}.
\end{equation}
Since $b_{10} = B_1a_1 + D_1a_{10} = B_1a_1 + D_1D_2D_3b_{10}$, we have:
\begin{equation}\label{p3_b_10}
    a_{10} = b_{30} = D_2D_3b_{10} = \frac{D_2D_3B_1}{1-D_1D_2D_3}a_1.
\end{equation}
Finally, using \eqref{p3_b_10}, we can express $b_1$ in terms of $a_1$ as follows:
\begin{equation}
    b_1 = A_1a_1 + B_1a_{10} = (A_1 + B^2_1D_2D_3/\Delta)a_1,
\end{equation}
where $\Delta = 1-D_1D_2D_3$.
\end{proof}
\end{theorem}

\begin{corollary}
If port 3 of the circulator is terminated with a tunable reactive impedance $Z_3$, the resulting two-port scattering matrix can be designed to have the following anti-diagonal scattering matrix:
\begin{equation}\label{eq_coro}
    \mathbf{\Phi} = \begin{bmatrix}
0 & e^{j\varphi_1} \\
e^{j\varphi_2} & 0
\end{bmatrix},
\end{equation}
for arbitrary $\varphi_1$ and $\varphi_2$.
\begin{proof}
The terminating impedance $Z_3$ results in $\frac{b_{30}}{a_{30}} = D_3 = \frac{Z_3-Z_0}{Z_3 + Z_0}$. We implement elements 1 and 2 with lossless transmission lines with a length corresponding to a phase shift of $e^{j\varphi_2/2}$. This is equivalent to letting $A_1 = A_2 = D_1 = D_2 = 0$ and $B_1=B_2 = B = e^{j\varphi_2/2}$. The resulting $2\times 2$ scattering matrix for the two element group becomes
\begin{equation}
    \mathbf{\Phi} = \begin{bmatrix}
0 & D_3 e^{j\varphi_2} \\
e^{j\varphi_2} & 0
\end{bmatrix}.
\end{equation}
\textcolor{black}{Since $Z_3 = jX_3$ is purely reactive, we have $D_3 = |D_3|e^{j\phi_3}$ with $|D_3|=1$ and $\phi_3 = -2\tan^{-1}(X_3/Z_0)$. Thus, $\phi_1 = \phi_2+\phi_3$ in \eqref{eq_coro} can be chosen arbitrarily by adjusting $Z_3$.
}
\end{proof}
\end{corollary}

\section{Signal Model and Non-Reciprocal Beamsteering}

\begin{figure}[t!]
    \begin{center}
        \includegraphics[scale=0.34]{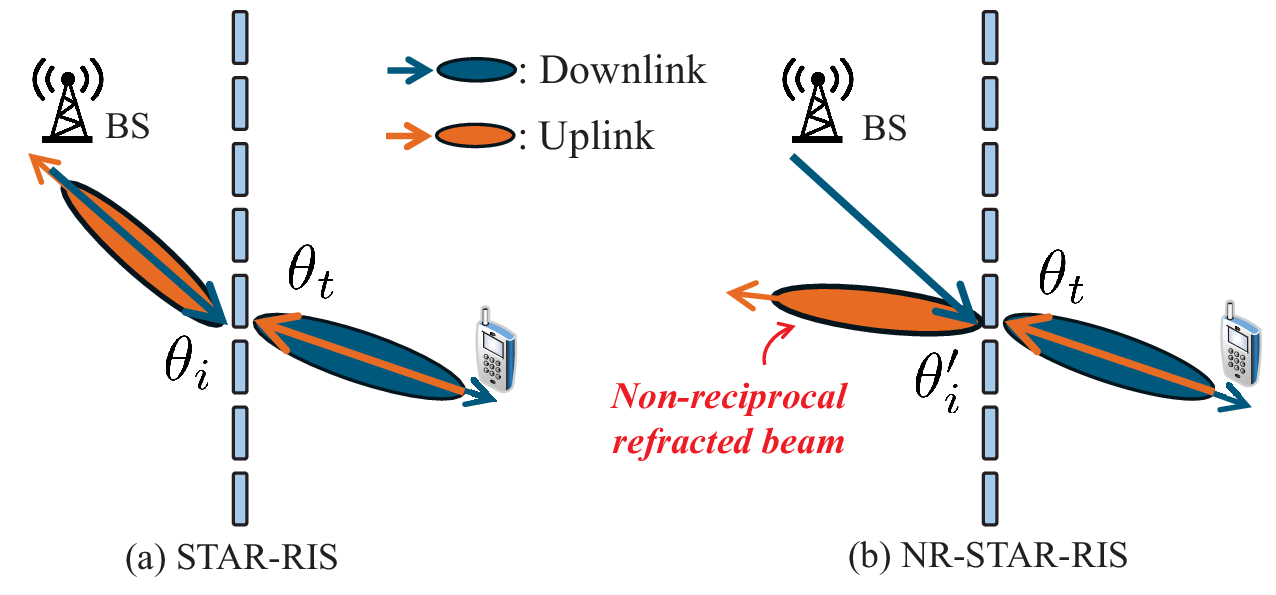}
        \vspace{-0.1 cm}
        \caption{Illustration of non-reciprocal beamsteering using NR-STAR-RISs.}
        \label{Fig_STAR}
    \end{center}\vspace{-0.4 cm}
\end{figure}

\begin{figure}[t!]
    \begin{center}
        \includegraphics[scale=0.32]{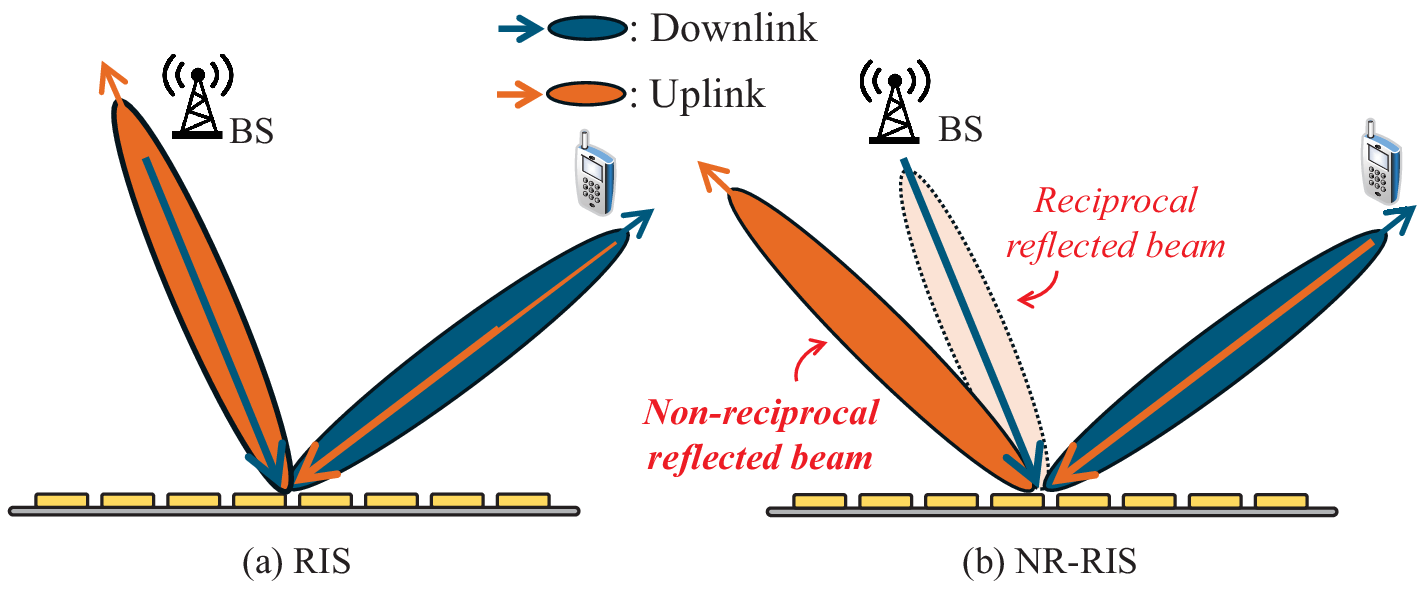}
        \vspace{-0.1 cm}
        \caption{Illustration of non-reciprocal beamsteering using NR-RISs.}
        \label{Fig_RIS}
    \end{center}\vspace{-0.4 cm}
\end{figure}

In this section we present signal models for an NR-RIS operating in either reflecting mode or simultaneously transmitting and reflecting (STAR) mode. Then, we demonstrate how to achieve non-reciprocal beamsteering. 
\subsection{Non-Reciprocal STAR Mode}
As illustrated in Fig.~\ref{Fig_STAR}, the basestation (BS) and the user are located on different sides of a STAR-RIS. Using the model proposed in the previous section, a STAR-element can be viewed as two reciprocal RIS elements connected through an NR circuit, with one on the BS side and the other on the user side. A downlink (DL) signal impinges only on the STAR-RIS via the left-side elements, while an uplink (UL) signal impinges only from the right. This creates a natural separation of DL and UL signals. In contrast, both the UL and DL signals are seen by all elements of a reflecting-only RIS, as illustrated in Fig.~\ref{Fig_RIS}. As a result, we will see that it is easier to design NR beamsteering using an NR-STAR-RIS.

We assume the STAR-RIS is constructed from $N/2$ two-element groups.
For the $n$-th element group, the scattering response is characterized as follows:
\begin{equation}\label{new_eq}
 \begin{bmatrix}
     b_{1,n} \\b_{2,n}
 \end{bmatrix}  = \begin{bmatrix}
\mathbf{\Phi}_{11,n} & \mathbf{\Phi}_{12,n} \\
\mathbf{\Phi}_{21,n} & \mathbf{\Phi}_{22,n}
\end{bmatrix} \begin{bmatrix}
     a_{1,n}\\a_{2,n}
 \end{bmatrix} \triangleq \ \mathbf{\Phi}_n \begin{bmatrix}
     a_{1,n}\\a_{2,n}
 \end{bmatrix}, 
\end{equation}
where $\mathbf{\Phi}_n$ is the scattering matrix for the $n$-th pair of STAR-RIS elements, the wavevariables $a_{1,n}, a_{2,n}$ respectively represent the incident signals on the left (1) and right (2) sides of the STAR-RIS, while $b_{1,n}, b_{2,n}$ are the corresponding reflected signals\footnote{We assume that all STAR-RIS element pairs are similarly oriented such that port 1 is facing the left and port 2 the right.}. 
The relationship governing the entire STAR-RIS is given by
\begin{equation}\label{signal_STAR}
    \mathbf{b} = \mathrm{blkdiag}\{\mathbf{\Phi}_1,\mathbf{\Phi}_2,\cdots,\mathbf{\Phi}_{N/2}\} \, \mathbf{a} \triangleq \bar{\mathbf{\Phi}} \mathbf{a},
\end{equation}
where \textcolor{black}{$\mathbf{b} = [b_{1,1} \: b_{2,1} \: \cdots \: b_{1,N/2} \: b_{2,N/2}]^\mathrm{T}$} is the transmitted/reflected signal vector and\\
$\mathbf{a} = [a_{1,1} \; a_{2,1} \: \cdots \: a_{1,N/2} \; a_{2,N/2}]^\mathrm{T}$ is the incident signal vector. The two transmission coefficients of a reciprocal STAR-RIS element are identical, i.e., \textcolor{black}{$\mathbf{\Phi}_{12,n}=\mathbf{\Phi}_{21,n}$}, and if the STAR-RIS is configured such that $\mathbf{\Phi}_{11,n}=\mathbf{\Phi}_{22,n}=0$, only reciprocal beamsteering can be achieved; i.e., the UL beam is refracted back to the direction of the BS downlink beam, as shown in Fig.~\ref{Fig_STAR}(a).

On the other hand, an NR-STAR-RIS configured with a non-symmetric anti-diagonal scattering matrix such as that in \textbf{Corollary 1} can achieve arbitrary non-reciprocal transmission coefficients. This implies that the UL and DL signals can be completely decoupled, i.e., 
$b_{1,n}=\mathbf{\Phi}_{12,n} a_{2,n}$ and $b_{2,n}=\mathbf{\Phi}_{21,n} a_{1,n}$.
In this case, we can easily find the phase-shifts that achieve non-reciprocal beamsteering in closed form. For example, consider a DL signal incident on side 1 from direction $\theta_i$, {i.e.}, $\mathbf{a}_1=[a_{1,1} \; a_{1,2} \; \cdots a_{1,N/2}]^\mathrm{T} = \mathbf{v}_{\frac{N}{2}}(\theta_i) s_d$, where for an NR-STAR-RIS configured as a uniform linear array (ULA) we have $\mathbf{v}_{\frac{N}{2}}(\theta) = [1, e^{-j\omega(\theta)}, \cdots,e^{-j(\frac{N}{2}-1)\omega(\theta)}]^\mathrm{T}$, $\omega(\theta)=2\pi d\sin(\theta)$ and $d$ is the element spacing in wavelengths. If we set $\mathbf{\Phi}_{21,n} = \exp \{ j2\pi(\sin\theta_i - \sin\theta_t)nd \}$, the NR-STAR-RIS will produce a refracted beam towards angle $\theta_t$ on side 2. On the other hand, if a user transmits its UL signal with incident angle $\theta_t$ from side 2, and the NR-STAR-RIS is configured such that $\mathbf{\Phi}_{12,n} = \exp \{ j2\pi(\sin\theta_t - \sin\theta'_i)nd \}$, then the UL refracted beam is steered towards $\theta'_i$ rather than~$\theta_i$.

\subsection{Non-Reciprocal Reflecting Mode}
%

Assume a reflection-only NR-RIS, and as before assume an incident DL signal $\mathbf{v}(\theta_i) s_d$ from direction $\theta_i$, where here we define $\mathbf{v}(\theta_i)=\mathbf{v}_N(\theta_i)$ to correspond to an $N$-element ULA. 
\textcolor{black}{
After reflection by the NR-RIS, the DL signal picked up by the antenna of a user at angle $\theta_l$ can be expressed as:}
\begin{equation}
    x(\theta_l) = \mathbf{v}^\mathrm{T}(\theta_l)\bar{\mathbf{\Phi}}\mathbf{v}(\theta_i)s_d.
\end{equation}
Now, consider an UL signal impinging on the NR-RIS from $\theta_l$. For an NR-RIS, the UL signal reflected by the surface in the direction of the DL source will in general be different from $x(\theta_l)$ since $\bar{\mathbf{\Phi}}\ne\bar{\mathbf{\Phi}}^\mathrm{T}$, and hence
\begin{equation}
\left(\mathbf{v}^\mathrm{T}(\theta_i)\bar{\mathbf{\Phi}}\mathbf{v}(\theta_l)\right)^T \ne \mathbf{v}^\mathrm{T}(\theta_l)\bar{\mathbf{\Phi}}\mathbf{v}(\theta_i) .
\end{equation}
While the UL beam will thus not point back at $\theta_i$, a natural question is can the UL beam be steered towards a specific desired direction other than $\theta_i$ by choosing a particular NR response $\bar{\mathbf{\Phi}}$. More generally, can we design $\bar{\mathbf{\Phi}}$ to simultaneously yield a desired DL beam response and a desired non-reciprocal UL beam response, as depicted in Fig.~\ref{Fig_RIS}? 
%

Here we address this question using the two-element group configuration discussed in \textbf{Corollary 1}, which yields a block diagonal scattering matrix composed of $2\times 2$ anti-diagonal blocks. Let $\theta_B$ and $\theta_U$ correspond to the directions from the NR-RIS to a basestation (B) and user (U), respectively, and let $P(\theta_l)$ and $P'(\theta_l)$ denote the desired DL and UL beampatterns, respectively, for a sampled set of angles $\theta_l, l=1,\cdots,L$. Using a beampattern matching criterion, we can formulate the NR-RIS configuration problem as follows:
 \begin{align} \label{mse1}
 \begin{split}
        &\underset{\begin{subarray}{c}
            \bar{\mathbf{\Phi}},\alpha,\alpha'
            \end{subarray}}{\mathrm{min}} \;\frac{1}{L} \sum\nolimits_{l=1}^L \Big| \big| 
            \mathbf{v}^\mathrm{T}(\theta_l)\bar{\mathbf{\Phi}}\mathbf{v}(\theta_B) \big|^2 - \alpha P(\theta_l) \Big|^2       \\
            & \qquad \qquad \qquad \qquad + \Big| \big| 
            \mathbf{v}^\mathrm{T}(\theta_{l})\bar{\mathbf{\Phi}}\mathbf{v}(\theta_U)  \big|^2 - \alpha' P'(\theta_l) \Big|^2 
        \end{split}\\
        \begin{split}
        & \quad\mathrm{s. t.} 
        \quad \bar{\mathbf{\Phi}} = \mathrm{blkdiag}\{\mathbf{\Phi}_1,\cdots,\mathbf{\Phi}_{\frac{N}{2}}\}, \ \mathbf{\Phi}_k = \begin{bmatrix}
0 & A_k \\
B_k & 0
\end{bmatrix}\\
        & \qquad |A_k| = |B_k| = 1,\ \forall k = 1,2, \ldots, N/2,
\end{split} \nonumber
\end{align}
where $\alpha\in\mathbb{R}$ and $\alpha'\in\mathbb{R}$ are scaling factors used to equalize the beam responses.
To transform \eqref{mse1} into a more easily solvable form, define $\bm{\phi}\in\mathbb{C}^N \triangleq [A_1,B_1, \cdots, A_{N/2},B_{N/2}]^T$ and $\bm{\Lambda}\triangleq \mathbf{I}_{N/2}\otimes\bm{\Gamma}$ with $\mathbf{\Gamma} = \bigg[\begin{matrix}
    0&1\\1&0
\end{matrix}\bigg]$. 
Then, problem \eqref{mse1} can be transformed as
 \begin{align} \label{mse2}
 \begin{split}
        \underset{\begin{subarray}{c}
            \bm{\phi},\alpha,\alpha'
            \end{subarray}}{\mathrm{min}} \quad &  \frac{1}{L} 
            \sum\nolimits_{l=1}^L \Big| \bm{\phi}^\mathrm{H}\mathbf{A}_{B,l}\bm{\phi} - \alpha P(\theta_l)
            \Big|^2 \\
            \quad & \qquad \qquad + \Big| \bm{\phi}^\mathrm{H}\mathbf{A}_{U,l}\bm{\phi} - \alpha' P'(\theta_l)
            \Big|^2 
        \\
        \mathrm{s. t.} \quad & | \bm{\phi}_i | = 1, \ \forall i=1,2,\ldots,N,
         \end{split}
        \end{align}
where 
$\mathbf{A}_{B,l} = \mathrm{diag}^H\{\mathbf{\Lambda}\mathbf{v}(\theta_B)\} \mathbf{v}^*(\theta_l) \mathbf{v}^\mathrm{T}(\theta_l) \mathrm{diag}\{\mathbf{\Lambda}\mathbf{v}(\theta_B)\}$, $(\cdot)^*$ denotes conjugation, and $\mathbf{A}_{U,l}$ is identical to $\mathbf{A}_{B,l}$ with $\theta_U$ in place of $\theta_B$. 
The objective is quadratic with respect to $\alpha$ or $\alpha'$, and both can be found in closed-form: 
\begin{equation}
    \alpha = \frac{\bm{\phi}^\mathrm{H}\sum_l P(\theta_l)\mathbf{A}_{B,l}\bm{\phi}}{\sum_l P^2(\theta_l)} \; , \; \alpha' = \frac{\bm{\phi}^\mathrm{H}\sum_{l} P(\theta_l)\mathbf{A}_{U,l}\bm{\phi}}{\sum_{l} P^2(\theta_{l})}.
\end{equation}
Substituting the optimal $\alpha$ and $\alpha'$ into \eqref{mse2}, the resulting optimization is quartic in $\bm{\phi}$ subject to element-wise unit-modulus constraints, which can be solved using manifold optimization~\cite{9534484}.

\begin{figure*}[t!]
\centering
\subfigure[DL beam pattern]{\label{n0}
\includegraphics[width= 2in]{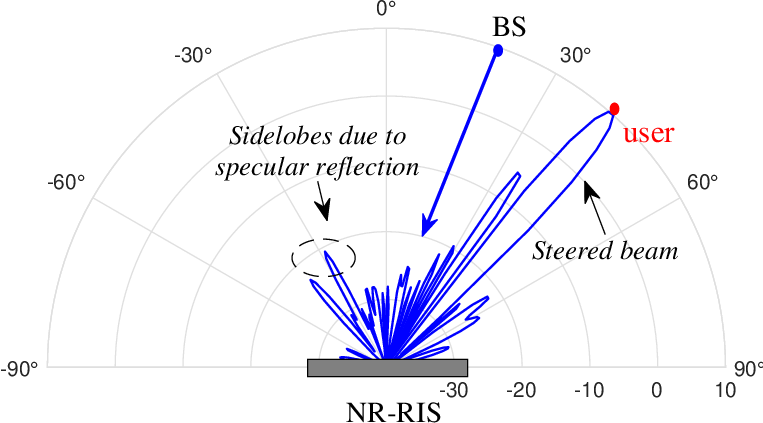}}
\subfigure[UL beam pattern]{\label{na}
\includegraphics[width= 2in]{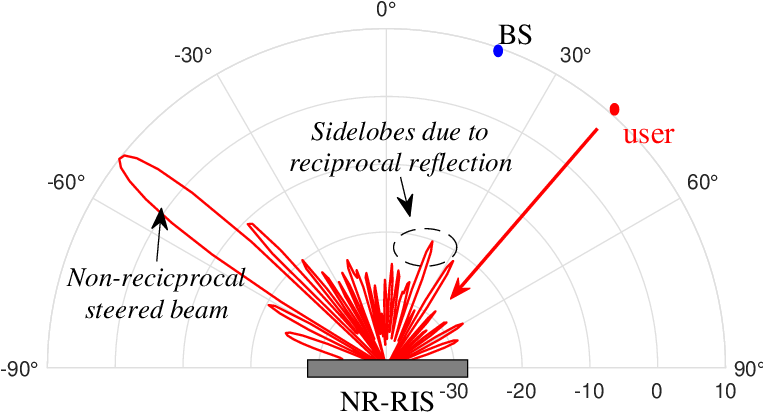}}
\subfigure[STAR-RIS beam pattern]{\label{nb}
\includegraphics[width= 2in]{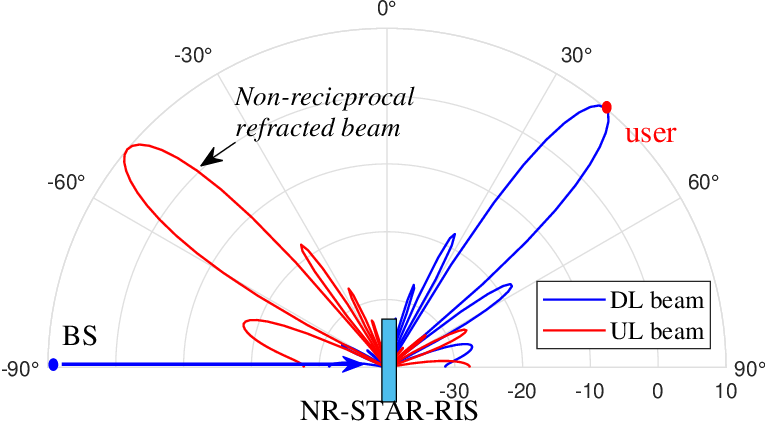}}
\vspace{-0.15 cm}
\caption{Radiation patterns of a three-element grouped NR-RIS and an NR-STAR-RIS in DL and UL.}\label{Fig_Duo}
\end{figure*}

\begin{figure}[t!]
    \begin{center}
        \includegraphics[scale=0.56]{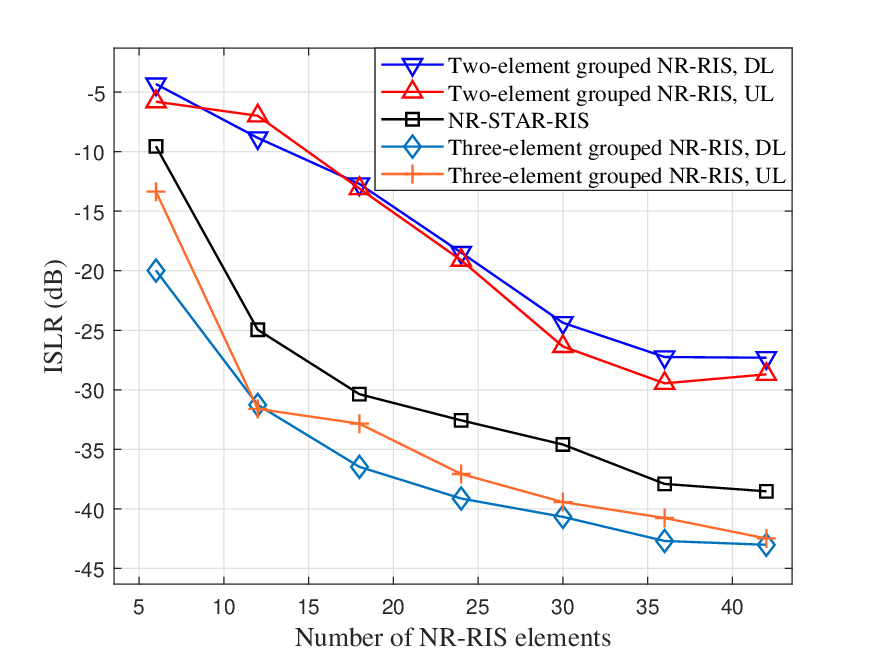}
        \vspace{-0.15 cm}
        \caption{ISLR v.s. the number of NR-RIS elements.}
        \label{Fig_num1}
    \end{center}
\end{figure}

\begin{figure}[t!]
    \begin{center}
        \includegraphics[scale=0.56]{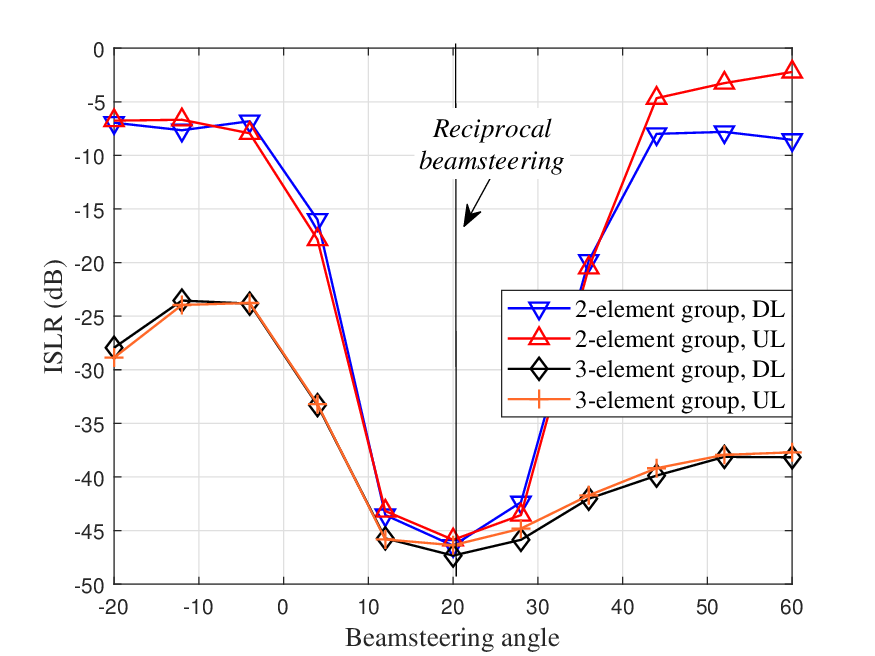}
        \vspace{-0.15 cm}
        \caption{ISLR v.s. beamsteering angle.}
        \label{Fig_num2}
    \end{center}
\end{figure}

\section{Numerical Results}\label{Num}

\begin{figure}[t!]\label{crack:fig}
\centering
\subfigure[MRT under CRACK.]{\label{crack1}
\includegraphics[width= 2.5in]{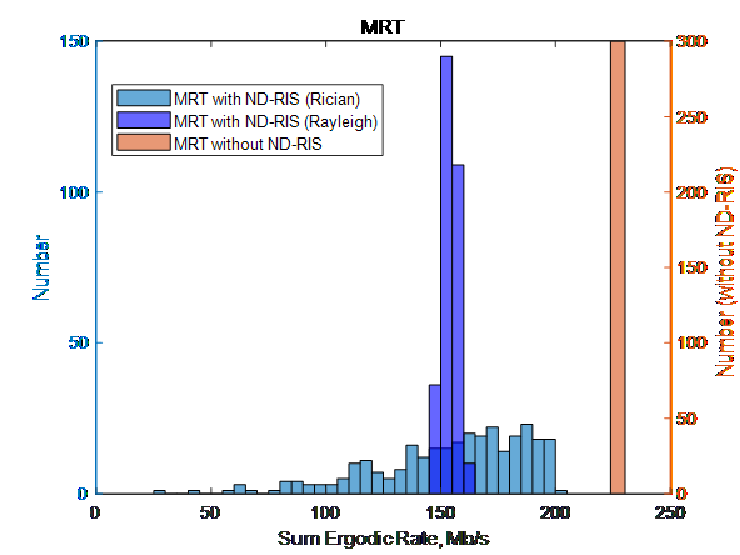}}
\subfigure[ZF under CRACK.]{\label{crack2}
\includegraphics[width= 2.5in]{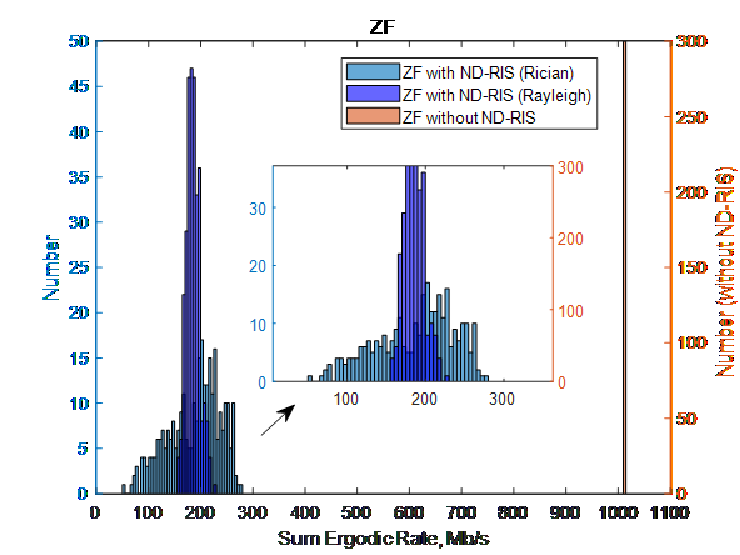}}

\caption{Histograms of ergodic rates with and without CRACK.}\label{Fig_Duo2}
\end{figure}

\subsection{Beamsteering}
%
In Fig.~\ref{Fig_Duo}, we show the radiation patterns for a linear NR-RIS array with a spacing of $d=1/2$ and a three-element grouped structure. Fig.~\ref{Fig_Duo}(a) shows a DL signal arriving at $\theta_B = 20^\circ$ and reflected towards a user at $\theta_U=40^\circ$, and Fig.~\ref{Fig_Duo}(b) shows an UL signal from $\theta_U$ being reflected at approximately $-50^\circ$, away from the DL source.  This demonstrates that strong non-reciprocity can be achieved using an NR-RIS with three-element NR groups.
Note that sidelobes in the reciprocal directions are present, but are suppressed by at least 30 dB.
In Fig.~\ref{Fig_Duo}(c), we plot the beam pattern achieved by an NR-STAR-RIS. For fairness, the NR-STAR-RIS has the same number of elements as the reflect-only NR-RIS, so only half as many elements are present on each side. As can be observed, the NR-STAR-RIS beams are wider but reciprocal sidelobes are essentially non-existant.

In Fig.~\ref{Fig_num1}, we plot the integrated sidelobe ratio (ISLR) versus the number of NR-RIS elements and show how the sidelobe power steadily decreases as the number of NR-RIS elements increases. 
\textcolor{black}{
Specifically, we compare the ISLRs for a two-element grouped NR-RIS implemented with circulators where the third port is reactively terminated (as described in {\bf Corollary 1}), a more general three-element grouped NR-RIS implemented with circulators (as discussed in {\bf Theorem 3}), and an NR-STAR-RIS.} 
The NR-RIS with general three-element circulator groups achieves the lowest ISLR, followed by the NR-STAR-RIS and the NR-RIS with two-element groups.
Fig.~\ref{Fig_num2} shows the ISLR versus the UL beamsteering angle for $\theta_B=20^\circ$ when the NR-RIS is optimized by solving problem \eqref{mse1}. We see that the sidelobes are better suppressed if the UL beamsteering angle is close to the reciprocal UL direction. The further the UL beam is steered away from $20^\circ$, the higher the sidelobe power. This behaviour demonstrates a trade-off between achieving strong non-reciprocity and suppressing sidelobes when designing the NR-RIS phase-shifts.

\subsection{Channel Reciprocity Attack}

In this case we consider the channel reciprocity attack (CRACK) presented in \cite{10445725}, but we use an NR-RIS composed of two-element groups based on the physically consistent model of {\bf Corollary~1} with $\phi_1=-\phi_2$ rather than the idealized ``permutation'' type non-symmetry assumed in \cite{9737373}. In CRACK, the NR-RIS creates different UL and DL channels that will degrade the performance of MIMO precoding methods predicated on reciprocity in time-division duplex systems. Fig.~\ref{Fig_Duo2} shows the significant degradation in ergodic sum rate that results when the NR-RIS is present for maximum ratio transmission (MRT) and zero-forcing (ZF) precoding. Compared with the results achieved in \cite{10445725} for the random permutation-type non-symmetry, the architecture based on the physically consistent model actually achieves a slightly increased performance degradation.

\section{Conclusions}
We have presented physically-consistent device and signal models for NR-RIS. NR phase-shift matrices were derived for two- and three-element grouped architectures connected with circulators. Based on the proposed models, we demonstrated that both NR-RIS and NR-STAR-RIS can achieve non-reciprocal beamsteering. We showed that the optimal NR-RIS configuration that produces a desired beampattern can be obtained by optimizing a quartic objective function subject to unit-modulus constraints. Numerical results confirmed that NR beamsteering can be achieved and sidelobe power is significantly reduced by increasing the number of NR-RIS elements. We also showed that our physically consistent NR configuration is also able to successfully implement a channel reciprocity attack.

\bibliographystyle{IEEEtran}
\bibliography{mybib}

\end{document}